\documentstyle[12pt]{article}
in

\def\beq{\begin{equation}}
\def\eeq{\end{equation}}
\def\bea{\begin{eqnarray}}
\def\eea{\end{eqnarray}}

\def\ba{\begin{array}}
\def\ea{\end{array}}
\def\ds{\displaystyle}

\def\,{\"{U}}
\def\6{\.{I}}

\begin{document}
\baselineskip 0.7cm
\title{ {Exact \Large{Solution of Schr\"{o}dinger equation with deformed Ring-Shaped
Potential}}}

\author{Metin Akta\c{s} and Ramazan Sever$\thanks{Corresponding author:
sever@metu.edu.tr}$\\
Department of Physics, Middle East Technical University \\
06531 Ankara, Turkey }

\date{\today}

\maketitle

\begin{abstract}

Exact solution of the  Schr\"{o}dinger equation with deformed ring shaped potential is obtained in the parabolic
and spherical coordinates. The Nikiforov-Uvarov method is used in
the solution. Eigenfunctions and corresponding energy eigenvalues are
calculated analytically. The agreement of our results is good.\\

\end{abstract}

\newpage
\section{Introduction}
\noindent
The exact solutions of the  Schr\"{o}dinger equation with some well-known
central and noncentral potentials is an important subject in
quantum mechanical problems. Such solutions are helpful for
checking and improving models and numerical methods besides of
understanding about the characteristics of a quantum system. 
The exact solutions of this equation are possible only 
for certain potentials such as Coulomb, Morse, P\"{o}schl-Teller
and harmonic oscillator etc. [1]. The other exactly solvable one is the 
deformed ring-shaped potential introduced by Hartmann [2]. This 
potential involves an attractive Coulomb potential with a repulsive 
inverse square potential one. In spherical coordinates it can be defined
as 
\beq \label{eq.1}
V(r,\theta)=\left[\frac{2}{r}-~q\delta\frac{a_{0}}{r^{2}sin^2\theta}\right]\delta\\
\sigma^{2}a_{0}\epsilon_{0}, 
\eeq 
\noindent where $a_{0}$ and
$\epsilon_{0}$ denote the Bohr radius and the ground state energy
of the hydrogen atom respectively. $\delta$ and $\sigma$ are
positive real parameters as well. Their range varies from 1 upto
10. This potential can be used in quantum chemistry and nuclear
physics to describe ring-shaped molecules like benzene and
interactions between deformed pair of nuclei. We point out that
the potential takes the form of the Coulomb potential in the
limiting case $\delta\sigma^{2}=Z$ and $q=0$ for hydrogen-like
atoms. The energy eigenvalues of the potential has been calculated
before by using some useful methods. For example, these are a
non-bijective canonical transformation, namely,
Kustaanheimo-Stiefel (KS) transformation, dynamical group method,
path integral and SUSYQM method etc. [2-17]. Moreover, this
potential can be defined as a Coulomb plus Aharonov-Bohm potential
by defining the parameters as
$-ee^{\prime}=2a_{0}\epsilon_{0}\delta\sigma^{2}$,
$-A/2\mu=q\epsilon_{0}a_{0}^{2}\delta^{2}\sigma^{2}$ and $B=0$.
[18, 19, 20].

 In the present work, we have obtained an exact solution
of the Schr\"{o}dinger equation  with the q-deformed ring shaped
potential by using the Nikiforov-Uvarov (NU) method in both
parabolic and spherical coordinates. The method is based on the
solution of differential equation transformed into the
hypergeometric type [21, 22].

The paper is organized as follows: In Section II we introduce the
Nikiforov-Uvarov method. In Section III we apply the method to
solve the Schr\"{o}dinger equation in both parabolic and spherical
coordinates respectively. In Section IV we present numerical results
for $Z=1$ and $q=0$ with the conclusion.

\section{The Nikiforov-Uvarov Method}
 The Nikiforov-Uvarov method first reduces the second order differential 
equations (ODEs) to
the hypergeometric type with an appropriate coordinate transformation
$x=x(s)$ as

\begin{equation} \label{eq.2}
\frac{d^{2}\Psi(s)}{ds^{2}}+\frac{\tilde{\tau}(s)}{\sigma(s)}
\frac{d\Psi(s)}{ds}+\frac{\tilde{\sigma}(s)}{{\sigma}^{2}(s)}\Psi(s)=0
\end{equation}

\noindent
where ${\sigma}(s)$ and ${\tilde{\sigma}(s)}$ are
polynomials with at most second degree, and
${\tilde{\tau}(s)}$ is a polynomial with at most first
degree [21, 22]. If we
take the following factorization

\begin{equation} \label{eq.3}
\Psi(s)=\phi(s)~y(s),
\end{equation}

\noindent
the Eq. (2) takes the form [22]

\begin{equation} \label{eq.4}
\sigma(s)\frac{d^{2}y(s)}{ds^{2}}+\tau(s)\frac{dy(s)}{ds}+\lambda y(s)=0,
\end{equation}

\noindent
where

\begin{equation} \label{eq.5}
\sigma(s)=\pi(s)\frac{d}{ds}(\ln\phi(s)),
\end{equation}
\noindent

\noindent
and
\begin{equation} \label{eq.6}
\tau(s)=\tilde{\tau}(s)+2\pi(s).
\end{equation}

\noindent Also, $\lambda$ is given

\begin{equation} \label{eq.7}
\lambda_{n}+n\tau^{'}+\frac{\left[n(n-1)\sigma^{\prime\prime}\right]}{2}=0,
\quad n=0,1,2,\ldots
\end{equation}

\smallskip
The energy eigenvalues can be calculated from the above equation. We
first have to determine $\pi(s)$ and $\lambda$ by defining

\begin{equation} \label{eq.8}
k=\lambda-\pi^{'}(s).
\end{equation}

\noindent
Solving the quadratic equation for $\pi(s)$ with the
Eq. (8), we get

\smallskip
\begin{equation} \label{eq.9}
\pi(s)=\left(\frac {\sigma^{\prime}-\tilde{\tau}}{2}\right)\pm
\sqrt{\left(\frac{\sigma^{\prime}-\tilde{\tau}}{2}\right)^{2}-
\tilde{\sigma}+k\sigma}.
\end{equation}

\noindent Here, $\pi(s)$ is a polynomial with the parameter $s$
and prime factors denote the differentials at first degree. The
determination of  $k$ is the essential point in the calculation of
$\pi(s)$. It is obtained by setting the discriminant of the square
root to zero [22]. Therefore, we obtain a general quadratic
equation for $k$.

\smallskip
The determination of the wave function is now in order. We consider
the Eq. (5) and the Rodrigues relation

\begin{equation} \label{eq.10}
y_{n}(s)=\frac{C_{n}}{\rho(s)}\frac{d^{n}}{ds^{n}}
\left[\sigma^{n}(s)~\rho(s)\right],
\end{equation}

\smallskip
\noindent where $C_{n}$ is normalization constant and the weight
function $\rho(s)$ satisfy the following relation

\begin{equation} \label{eq.11}
\frac{d}{ds}\left[\sigma(s)~\rho(s)\right]=\tau(s)~\rho(s).
\end{equation}

\smallskip
\noindent
The Eq. (10) refers to the classical orthogonal
polynomials that have many important properties especially
orthogonality relation can be defined as

\smallskip
\begin{equation} \label{eq.12}
\int_{a}^{b}y_{n}(s)~ y_{m}(s) \rho(s)~ds=0 ,\quad m\neq n.
\end{equation}

\section{Calculations}

The Schr\"{o}dinger equation in spherical coordinates becomes

\begin{equation} \label{eq.13}
\nabla^{2}\Psi+\frac{2m}{{\hbar}^{2}}\left[E-V(r, \theta)\right]\Psi=0.
\end{equation}

\noindent
We are first going to study for solution of the problem using
parabolic coordinates.

\subsection{Parabolic coordinates}
\noindent
One can write the second type parabolic coordinates as [3, 4, 23]
\begin{equation} \label{eq.14}
x= \xi\eta~cos\varphi
 ,\quad y= \xi\eta~sin\varphi \quad and \quad
z=\frac{1}{2}\left(\eta^{2}-\xi^{2}\right),
\end{equation}

\noindent and

\begin{equation} \label{eq.15}
\xi~\eta=r~sin \theta \quad and \quad
r=\frac{1}{2}\left(\eta^{2}+\xi^{2}\right).
\end{equation}

\noindent If we write trial wave function in the following form [4]
\begin{equation} \label{eq.16}
\Psi(\xi, \eta,
\varphi)=\frac{1}{\sqrt{\xi\eta}}~u(\xi)~v(\eta)~e^{im^{\prime}\varphi},
\end{equation}

\noindent one can get two-coupled differential equations

\begin{equation} \label{eq.17}
\frac{d^{2}u}{d\xi^{2}}-\frac{(\Upsilon^{2}-\frac{1}{4})}{\xi^{2}}u+
(\frac{2mE}{\hbar^{2}})\xi^{2}u-(\frac{2m}{\hbar^{2}})\mu_{1}u=0,
\end{equation}

\noindent and

\begin{equation} \label{eq.18}
\frac{d^{2}v}{d\eta^{2}}-\frac{(\Upsilon^{2}-\frac{1}{4})}{\eta^{2}}v+
(\frac{2mE}{\hbar^{2}})\eta^{2}v-(\frac{2m}{\hbar^{2}})\mu_{2}v=0,
\end{equation}

\bigskip
\noindent where
$\Upsilon=\sqrt{m^{{\prime}^2}+~q\delta^{2}\sigma^{2}}$ and ${\ds
\mu_{1}=\mu_{2}=2\sigma^{2}~\delta~a_{0}~\epsilon_{0}}$. We will first
solve the Eq. (17) and then easily get the other one.

\smallskip
\noindent By using the transformation $s=\xi^{2}$, the Eq. (17) is
therefore transformed into the equation of hypergeometric
type. Hence, we have

\begin{equation} \label{eq.19}
u^{\prime\prime}(s)+\frac{1}{2s} u^{\prime}(s)
+\frac{1}{4s^{2}}\left[-\varepsilon^{2}s^{2}-\alpha_{1}^{2}s-\beta^{2}\\
\right]u(s)=0,
\end{equation}

\bigskip
\noindent where ${\ds\varepsilon^{2}=\frac{2mE}{\hbar^{2}}}$,\quad
${\ds \alpha_{1}^{2}=\frac{2m}{\hbar^{2}}\mu_{1}}$\quad and \quad
${\ds\beta^{2}=(\Upsilon^{2}-\frac{1}{4}})$.

\bigskip
\noindent Comparing the Eq. (19) with the Eq. (2), we get

\smallskip
\begin{equation} \label{eq.20}
\sigma(s)=2s,\quad\tilde{\tau}(s)=1 \quad
and\quad\tilde{\sigma}(s)=
\left(-\varepsilon^{2}s^{2}-\alpha_{1}^{2}s-\beta^{2}\right).
\end{equation}

\noindent Substituting these into the Eq. (9), we write

\smallskip
\begin{equation} \label{eq.21}
\pi(s)=\frac{1}{2}\pm\frac{1}{2}\sqrt{4\varepsilon^{2}s^{2}
+(8k+4\alpha_{1}^{2})s+4\beta^{2}}.
\end{equation}

\noindent The constant $k$ can be determined from the condition that the
discriminant of the square root must be zero, so that

\begin{equation} \label{eq.22}
k_{1,2}=-\frac{1}{2}\alpha_{1}^{2}\pm
\varepsilon\beta.
\end{equation}

\noindent Hence the final result for Eq. (21) can be written as

\begin{eqnarray} \label{eq.23}
\pi(s)=\frac{1}{2}\pm
\left\{
\begin{array}{ll}
(\varepsilon s-\beta), ~~~~&\mbox{$for ~~~~
k=-\frac{1}{2}\alpha_{1}^{2}-\varepsilon\beta$}\\[0.5cm]
(\varepsilon s+\beta), ~~~~&\mbox{$for ~~~~
k=-\frac{1}{2}\alpha_{1}^{2}+\varepsilon\beta$}.
\end{array}
\right.
\end{eqnarray}

\noindent A proper value for $\pi(s)$ is chosen, so that the function

\begin{equation} \label{eq.24}
\tau(s)=2(1+\beta)-2\varepsilon s,
\end{equation}
must have a negative derivative [22]. From the Eq. (7) we can obtain

\begin{eqnarray} \label{eq.25}
\lambda_{n}&=&-\frac{1}{2}\alpha_{1}^{2}-\varepsilon-\beta\varepsilon \nonumber \\
       &=&2n \varepsilon.
\end{eqnarray}

\noindent Following the same procedure again one gets for Eq. (18)
as

\begin{eqnarray} \label{eq.26}
\lambda_{n^{\prime}}&=&-\frac{1}{2}\alpha_{2}^{2}-\varepsilon
-\beta\varepsilon \nonumber \\
                   &=&2n^{\prime} \varepsilon.
\end{eqnarray}

 By combining each side of the Eqs. (25) and (26) we
obtain energy eigenvalues

\smallskip
\begin{equation} \label{eq.27}
E_{n,n^{\prime}}=-\left[\frac{\delta^{2}~\sigma^{4}}{\left(n+n^{\prime}+1\\
+\beta\right)^{2}}\right]\epsilon_{0}.\\
\quad\quad n, n^{\prime}=0,1,2\ldots
\end{equation}

\smallskip
\noindent This solution is identical for 
$\beta\simeq\Upsilon^{2}$ with the ones obtained
before [3, 4, 6, 7, 8, 9, 13].

 Now, we are going to determine the wave function. Considering the
Eq. (3) and using the Eq. (5) we get

\begin{equation} \label{eq.28}
\phi(s)=s^{\nu/4}~e^{-\frac{\varepsilon}{2} s},
\end{equation}

\noindent where $\nu=1+2\beta$.

\noindent From the equations Eqs. (11) and (10), we obtain

\smallskip
\begin{equation} \label{eq.29}
y_{n}(s)=\frac{C_{n}}{\rho(s)}\frac{d^{n}}{ds^{n}}\left[s^{n}~\rho(s)\right],
\end{equation}

\noindent with ${\ds\rho(s)=s^{(\nu-1)/2}~e^{-\varepsilon s}}$.
The Eq. (33) stands for the associated Laguerre polynomials. That
is

\begin{equation} \label{eq.30}
y_{n}(s)\equiv L_{n}^{\beta}(s),
\end{equation}

\noindent Hence we have found the wave function that
belongs to the Eq. (17) as

\smallskip
\begin{equation} \label{eq.31}
u_{n}(\xi)=C_{n}~s^{\nu/4}~e^{-\frac{\varepsilon}{2} s}~L_{n}^{\beta}(s),
\end{equation}

\smallskip
\noindent with ${\ds s=\xi^{2}}$. Similarly, we can also write the wave
function for Eq. (18)

\smallskip
\begin{equation} \label{eq.32}
v_{n}(\eta)=C_{n^{\prime}}~s^{\nu/4}~e^{-\frac{\varepsilon}{2}
s}~L_{n^{\prime}}^{\beta}(s),
\end{equation}

\noindent with ${\ds s=\eta^{2}}$. Therefore, the total wave
function takes 
\beq \label{eq.33} 
\Psi_{n, n^{\prime},m^{\prime}}(\xi,\eta,\varphi)=\frac{1}{\sqrt{\xi~\eta}}
~C_{{n},n^{\prime}}~s^{\nu/2}~e^{-\varepsilon
s}~L_{n}^{\beta}(s)~L_{n^{\prime}}^{\beta}(s)~e^{im^{\prime}\varphi},
\eeq 
where the normalization constant  $C_{{n},n^{\prime}}$  can
be found from the Eq. (12) as

\smallskip
\begin{equation} \label{eq.34}
C_{n,n^{\prime}}=\sqrt{\frac{4(n!)(n^{\prime})!}{(n+\beta)!~(n^{\prime}+
\beta)!}},\quad\quad n,~n^{\prime}=0,1,2\ldots
\end{equation}

One can easily see that in the case of ${\ds r
sin\theta=\xi\eta}$, the problem reduces to harmonic oscillator
plus inverse square potential case. The latter, we have studied
that this problem also reduces to that of molecular Kratzer
potential like (Coulomb plus inverse square). 

\subsection{Spherical coordinates}
\noindent Considering the Eq. (13) we write the total wave
function as
\begin{equation} \label{eq.35}
\Psi(r, \theta, \varphi)=\frac{U(r)}{r}~\Theta(\theta)~\Phi(\varphi),
\end{equation}

\noindent with the well-known azimuthal angle solution

\begin{equation} \label{eq.36}
\Phi(\varphi)=\frac{1}{\sqrt{2\pi}}~e^{im\varphi}, \quad m=0,\pm 1,\pm
2,\ldots
\end{equation}

\noindent Thus we get as

\begin{equation} \label{eq.37}
\frac{1}{sin\theta}\frac{d}{d\theta}\left(sin\theta \\
\frac{d\Theta}{d\theta}\right)+\left(\kappa-\frac{(m^{2}+b^{\prime})}
{sin^{2}\theta}\right)\Theta=0,
\end{equation}
\smallskip
\noindent and

\begin{equation} \label{eq.38}
\frac{d^{2}U}{dr^{2}}+\frac{2}{\gamma}\left(-E^{\prime}-\frac{a^{\prime}}
{r}-\frac{\kappa}{r^{2}}\right)U=0,
\end{equation}

\smallskip
\noindent where ${\ds E^{\prime}=\frac{2mE}{\hbar^{2}}}$, ${\ds
a^{\prime}=\frac{2ma}{\hbar^{2}}}$ and ${\ds
b^{\prime}=\frac{2mb}{\hbar^{2}}}$, $\kappa$ and $m^{2}$ are also
seperation constants.

\smallskip
 Using the NU-method, we are going to solve them. By defining
$m^{\prime}=\sqrt{m^{2}+b}$ 
in Eq. (37) and taking $x=cos\theta$, it will have a form of
hypergeometric type

\beq \label{eq.39}
\frac{d^{2}\Theta}{dx^{2}}-\frac{2x}{(1-x^{2})}\frac{d\Theta}{dx} \\
+\frac{1}{(1-x^{2})^{2}}\left[\kappa(1-x^{2}) \\
-{m^{\prime}}^{2}\right]\Theta(x)=0.
\eeq

\smallskip
\noindent Comparing it with the Eq. (2), we get

\begin{equation} \label{eq.40}
\sigma(x)=x,\quad\tilde{\tau}(x)=-2x \quad
and\quad\tilde{\sigma}(x)=\kappa(1-x^{2})-{m^{\prime}}^{2}.
\end{equation}

\noindent Substituting these into the Eq. (9), we get

\beq \label{eq.41}
\pi(x)=\pm\sqrt{-(k+\kappa)(1-x^{2})+{m^{\prime}}^{2}}.
\eeq

\noindent The constant k is determined from the condition that the
discriminant of the square root must be zero. Thus, we find

\begin{eqnarray} \label{eq.42}
\pi(x)=
\left\{
\begin{array}{ll}
\pm~ m^{\prime}, ~~~~&\mbox{$for ~~~~
k=\kappa$} \\[0.5cm]
\pm~m^{\prime}~x, ~~~~&\mbox{$for ~~~~
k=\kappa-{m^{\prime}}^{2}$}.
\end{array}
\right.
\end{eqnarray}

\noindent A proper value for $\pi(x)$ can be chosen, so that the function

\begin{equation} \label{eq.43}
\tau(x)=-2(m^{\prime}+1)~x,
\end{equation}
has a negative derivative. From the Eq. (7) we can obtain

\begin{eqnarray} \label{eq.44}
\lambda_{n}&=&\kappa-m^{\prime}(m^{\prime}+1) \nonumber \\
       &=&2n~(m^{\prime}+1)+n~(n-1).
\end{eqnarray}

\noindent Solving for $\kappa$, we have

\beq \label{eq.45}
\kappa=\kappa_{n}=\ell^{\prime}(\ell^{\prime}+1),
\eeq

\noindent where $\ell^{\prime}=n+m^{\prime}$.

 Now we are going to determine the wave function. From the Eqs. (11) and (10), 
we can write

\smallskip

\begin{equation} \label{eq.46}
y_{n}(x)=\frac{B_{n}}{\rho(x)}\frac{d^{n}}{dx^{n}} \\
\left[(1-x^{\prime})^{n+m^{\prime}}\right],
\end{equation}

\noindent with $\rho(x)=(1-x^{2})^{m^{\prime}}$. Therefore the Eq.
(46) stands for Jacobi polynomial as

\beq \label{eq.47}
y_{n}\equiv~P_{n}^{(m^{\prime}, m^{\prime})}(x),
\eeq

\noindent where $n=l^{\prime}-m^{\prime}$. The wave function becomes

\begin{eqnarray} \label{eq.48}
\Theta(x)&=&\Theta_{l^{\prime}, m^{\prime}} \nonumber \\
         &=&C_{l^{\prime}, m^{\prime}}~(1-x^{2})^{m^{\prime}/2}
           ~P_{l^{\prime}-m^{\prime}}^{(m^{\prime}, m^{\prime})}(x),
\end{eqnarray}

\noindent with $x=cos\theta~(x\in[-1,1])$. Using the Eq. (12), we get

\beq \label{eq.49}
C_{\ell^{\prime},m^{\prime}}=\frac{1}{2^{m^{\prime}}(\ell^{\prime}+1)} 
\sqrt{\frac{2\ell^{\prime}+1}{2}(\ell^{\prime}-m^{\prime})!(\ell^{\prime}\\
+m^{\prime})!}.
\eeq

\smallskip
 Let us now consider the Eq. (38)

\beq \label{eq.50}
u^{\prime\prime}(r)+\frac{1}{r^{2}}\left[-E^{\prime}r^{2}\\
-a^{\prime}r-\kappa\right]u(r)=0,
\eeq

\noindent Comparing this equation with the Eq. (2), we obtain

\begin{equation} \label{eq.51}
\sigma(r)=r,\quad\tilde{\tau}(r)=0 \quad
and\quad\tilde{\sigma}(r)=-E^{\prime}r^{2}-a^{\prime}r-\kappa.
\end{equation}

\noindent If we insert these into the Eq. (9), one gets

\beq \label{eq.52}
\pi(r)=\frac{1}{2}\pm\sqrt{4E^{\prime}r^{2}+4(k+a^{\prime})r \\
+(1+4\kappa)}.
\eeq

\noindent We can determine the constant $k$ by using the condition
that discriminant of the square root is zero, that is

\beq \label{eq.53}
k_{1,2}=-a^{\prime}\pm\sqrt{E^{\prime}~(1+4\kappa)}.
\eeq

\noindent Hence the final form of the Eq. (52) for each value of $k$
becomes

\begin{eqnarray} \label{eq.54}
\pi(r)=\frac{1}{2}\pm\frac{1}{2}
\left\{
\begin{array}{ll}
[2\sqrt{E^{\prime}}r+\sqrt{1+4\kappa}],
~~~~&\mbox{$for ~~~~k=-a^{\prime}+\sqrt{E^{\prime}(1+4\kappa)}$}\\[0.5cm]
[2\sqrt{E^{\prime}}r-\sqrt{1+4\kappa}],
~~~~&\mbox{$for ~~~~k=-a^{\prime}-\sqrt{E^{\prime}(1+4\kappa)}$}.
\end{array}
\right.
\end{eqnarray}

\noindent A proper value for $\pi(r)$ is taken, so that the function

\beq \label{eq.55}
\tau(r)=(1+\sqrt{1+4\kappa})-2\sqrt{E^{\prime}}r,
\eeq

\noindent has a negative derivative. From the Eq. (7), we
can write

\begin{eqnarray} \label{eq.56}
\lambda_{n}&=&-a^{\prime}-\sqrt{E^{\prime}(1+4\kappa)}-\sqrt{E^{\prime}}\nonumber\\
           &=&2n~\sqrt{E^{\prime}}.
\end{eqnarray}

\noindent Therefore, it gives us the energy eigenvalues of the
radial equation with the deformed ring-shaped potential

\beq \label{eq.57}
E=\left[-\left(\frac{\delta^{2}~\sigma^{4}}{n_{r}
+\ell^{\prime}+1}\right)^{2}\epsilon_{0}\right],\quad n_{r}=0,1,2,\ldots
\eeq

\noindent where $n_{r}$ denotes the radial quantum number which belongs to
the Eq. (38).

\smallskip
 To determine the wave function, we consider the
Eqs. (3) and (5) for obtaining

\beq \label{eq.58}
\phi(r)=~e^{-\sqrt{E^{\prime}}~r}~r^{(\nu-1)/2},
\eeq

\noindent where $\nu=1+2\sqrt{1+4\kappa}$. Thus from the Eqs. (11) and
(10) we have

\begin{equation} \label{eq.59}
y_{n}(r)=\frac{B_{n}}{\rho(r)}\frac{d^{n}}{dr^{n}} \\
\left[\sigma^{n(r)}~\rho(r)\right],
\end{equation}

\noindent with ${\ds
\rho(r)=e^{-\sqrt{E^{\prime}}r}~r^{(1-\nu)/2}}$. The Eq. (59)
stands for associated Laguerre polynomials, that is

\beq \label{eq.60}
y_{n}(r)=L_{n}^{\bar k}(r),
\eeq

\noindent where $\bar k=(\nu-1)/2$. The radial part wave function is
written as

\beq \label{eq.61}
U_{n}(r)=C_{n}~e^{-\sqrt{E^{\prime}}r}~r^{\bar k}~L_{n}^{\bar k}(r).
\eeq

\noindent By using the orthogonality condition, we can determine
the coefficient as

\beq \label{eq.62}
C_{n,\bar k}=\sqrt{\frac{n!}{2(n+\bar k)(n+\bar k)!}},
\eeq

\noindent with $\kappa=\ell^{\prime}(\ell^{\prime}+1)$. Hence,  the total
wave function takes the form

\begin{eqnarray} \label{eq.63}
\nonumber \Psi(r,\theta, \varphi)
&=&\frac{1}{\sqrt{2\pi}}~C_{n,
\bar k}~C_{\ell^{\prime}, m^{\prime}}
\left[
e^{-\sqrt{E^{\prime}}r}~r^{\bar k}~(Sin\theta)^{m^{\prime}}\right.\nonumber\\
\smallskip
&\times&\left.~P_{n}^{(m^{\prime}, m^{\prime})}
(cos\theta)~L_{n}^{\bar k}(r)~e^{im\varphi}
\right].
\end{eqnarray}

\section{Conclusion and Remarks}
 We have obtained the exact eigenfunctions and corresponding 
energy eigenvalues of the
Schr\"{o}dinger equation with the deformed ring-shaped potential in
both second type parabolic and also spherical coordinates by using the
Nikiforov-Uvarov method. At first our
problem reduces to the harmonic oscillator plus inverse square potential, 
it also reduces to the problem that molecular Kratzer (Coulomb plus
inverse square) one in second case. Results obtained in two different 
coordinate systems 
are identical by following the conditions  $\beta\simeq\Upsilon^{2}$ in
Eq. (27) and  $\ell^{\prime}=n+m^{\prime}$ in Eq. (57). Some numerical 
values of energy  for a hydrogen-like atom due 
to the attractive Coulomb potential are presented in tabular form.
 The total wave functions, in both coordinates, are physical. They behaves
asymptotically. The agreement of our analytic and numerical results is 
good.


\newpage

{\bf Table I:}~~{\small Energy levels for a hydrogen-like atom with
$\delta\sigma^{2}=Z$ and $q=0$.}\\

\begin{tabular}{ccccc}\hline\hline
\\ $\mathbf{m} 
$ & $
\hspace*{0.5cm}
\mathbf{n+n^{\prime}}
\hspace*{0.5cm}
$ & $
\hspace*{0.5cm}
\mathbf{E_{n,~n^{\prime}}~(Our~work)} 
\hspace*{0.5cm}
$ & $
\hspace*{0.5cm}
\mathbf{E~[18]}
\hspace*{0.5cm}
$ & $
\mathbf{\bar n}$ \\[0.5cm] \hline

0~~&~~0~~&~~-13.605820~~&~~-13.60582~~&~~1 \\[0.2cm]\hline      
1~~&~0                              \\ [0.2cm]      
0~~&~~1~~&~~\raisebox{3mm}{-3.401455}~~&~~\raisebox{3mm}{-3.40145}~~&~~\raisebox{3mm}{2}\\[0.2cm]
\hline

2~~&~0                              \\[0.2cm]
1~~&~~1~~&~~-1.511757~~&~~-1.51176~~&~~3 \\[0.2cm]             
0~~&~2                              \\[0.2cm]\hline

3~~&~0                              \\[0.2cm]
2~~&~1                              \\[0.2cm]
1~~&~~2~~&~~\raisebox{2mm}{-0.850363}~~&~~\raisebox{2mm}{-0.85036}~~&~~\raisebox{2mm}{4}\\[0.2cm]
0~~&~2                              \\[0.2cm]\hline

4~~&~0                                 \\[0.2cm]
3~~&~1                                 \\[0.2cm]
2~~&~~2~~&~~-0.544232~~&~~-0.54423~~&~~5    \\[0.2cm]
1~~&~3                                 \\[0.2cm]
0~~&~4                                 \\[0.2cm]\hline

5~~&~0                                 \\[0.2cm]
4~~&~1                                 \\[0.2cm]
3~~&~2                                 \\[0.2cm]
2~~&~~3~~&~~\raisebox{2mm}{-0.377939}~~&~~\raisebox{2mm}~~{}~~&~~\raisebox{2mm}{6}\\[0.2cm]
1~~&~4                                 \\[0.2cm]
0~~&~5                                 \\[0.2cm]\hline
\end{tabular}

\end{document}